\documentclass[sigconf]{acmart}

\usepackage{booktabs} 
\usepackage{subfig}
\usepackage{bm}
\usepackage{balance}
\usepackage[ruled]{algorithm2e}
\usepackage{algorithmic}
\usepackage{multirow}
\usepackage{tikz}
\usetikzlibrary{positioning}

\setcopyright{rightsretained}
\copyrightyear{2019}
\acmYear{2019}
\acmConference[ICTIR '19]{The 2019 ACM SIGIR International Conference on the Theory of Information Retrieval}{October 2--5, 2019}{Santa Clara, CA, USA} 
\acmBooktitle{The 2019 ACM SIGIR International Conference on the Theory of Information Retrieval (ICTIR '19), October 2--5, 2019, Santa Clara, CA, USA} \acmDOI{10.1145/3341981.3344218}
\acmISBN{978-1-4503-6881-0/19/10}

\begin{document}
\fancyhead{}

\title{Learning Groupwise Multivariate Scoring Functions Using Deep Neural Networks}

\author{Qingyao Ai}
\affiliation{%
  \institution{CICS, UMass Amherst}
  \city{Amherst} 
  \state{MA} 
  \country{USA}
}
\email{aiqy@cs.umass.edu}

\author{Xuanhui Wang}
\affiliation{%
  \institution{Google Research}
    \city{Mountain View} 
	\state{CA} 
	\country{USA}
}
\email{xuanhui@google.com}

\author{Sebastian Bruch}
\affiliation{%
  \institution{Google Research}
    \city{Mountain View} 
	\state{CA} 
	\country{USA}
}
\email{bruch@google.com}

\author{Nadav Golbandi}
\affiliation{%
  \institution{Google Research}
    \city{Mountain View} 
	\state{CA} 
	\country{USA}
}
\email{nadavg@google.com}

\author{Michael	Bendersky}
\affiliation{%
  \institution{Google Research}
    \city{Mountain View} 
	\state{CA} 
	\country{USA}
}
\email{bemike@google.com}

\author{Marc Najork}
\affiliation{%
  \institution{Google Research}
    \city{Mountain View} 
	\state{CA} 
	\country{USA}
}
\email{najork@google.com}

\renewcommand{\shortauthors}{Ai et al.}

\begin{abstract}

While in a classification or a regression setting a label or a value is assigned to each individual document, in a ranking setting we determine the relevance ordering of the entire input document list. This difference leads to the notion of relative relevance between documents in ranking. The majority of the existing learning-to-rank algorithms model such relativity at the loss level using pairwise or listwise loss functions. However, they are restricted to \emph{univariate scoring functions}, i.e., the relevance score of a document is computed based on the document itself, regardless of other documents in the list. To overcome this limitation, we propose a new framework for \emph{multivariate scoring functions}, in which the relevance score of a document is determined jointly by multiple documents in the list. We refer to this framework as GSFs---groupwise scoring functions. We learn GSFs with a deep neural network architecture, and demonstrate that several representative learning-to-rank algorithms can be modeled as special cases in our framework. We conduct evaluation using click logs from one of the largest commercial email search engines, as well as a public benchmark dataset. In both cases, GSFs lead to significant performance improvements, especially in the presence of sparse textual features.

\end{abstract}

%
%



\maketitle

\section{Introduction}\label{sec:intro}

Unlike in classification or regression, the main goal of a ranking problem is not to assign a label or a value to individual items,
but, given a list of items, to produce an ordering of the items in that list in such a way that the utility of the entire list is maximized.
In other words, in ranking we are more concerned with the relative ordering of the relevance of items---for some notion of relevance---than their absolute magnitudes.

Modeling relativity in ranking has been extensively studied in the past, especially in the context of learning-to-rank~\cite{liu2009learning}. 
Learning-to-rank aims to learn a scoring function that maps feature vectors to real-valued scores in a supervised setting.
Scores computed by such a function induce an ordering of items in the list.
The majority of existing learning-to-rank algorithms learn a parameterized function by optimizing a loss that acts on pairs of items (\emph{pairwise}) or a list of items (\emph{listwise})~\cite{burges2005learning, quoc2007learning, cao2007learning, Jun+Hang:2007}.
The idea is that such loss functions guide the learning algorithm to optimize preferences between pairs of items or to maximize a ranking metric such as NDCG~\cite{Joachims:2002, Taylor+al:2008, burges2010ranknet},
thereby indirectly modeling relative relevance.

Though effective, most existing learning-to-rank frameworks are restricted to the paradigm of \emph{univariate} scoring functions: the relevance of an item is computed independently of other items in the list.
This setting could prove sub-optimal for ranking problems for two main reasons.
First, univariate functions have limited power to model cross-item comparison. 
Consider an \emph{ad hoc} document retrieval scenario where a user is searching for the name of an artist. 
If all the results returned by the query (e.g.,  ``calvin harris'') are recent, the user may be interested in the latest news or tour information. 
If, on the other hand, most of the query results are older (e.g., ``frank sinatra''), it is more likely that the user seeks information on artist discography or biography. Thus, the relevance of each document depends on the distribution of the whole list.
Second, user interaction with search results shows a strong tendency to \emph{compare} items.
Prior research suggests that preference judgments by comparing a pair of documents are faster to obtain, and are more consistent than absolute ratings~\cite{ye2013combining}. 
Moreover, it has been shown that better predictive capability is achieved when user actions are modeled in a relative fashion (e.g., SkipAbove)~\cite{Joachims+al:2005, Borisov+al:2016}.
These studies indicate that users compare a document with its surrounding documents prior to a click,
and that a ranking model that uses the direct comparison mechanism can be more effective, as it mimics user behavior more closely.

Given the above arguments, we hypothesize that the relevance score of an item should be computed by comparison with other items in the list at the feature level.
Specifically, we explore a general setting of \emph{multivariate} scoring functions for learning-to-rank. 
In its general form, a multivariate scoring function $f: \mathcal{X}^n \rightarrow \mathbb{R}^n$, where $\mathcal{X}$ is the universe of all items, takes a vector of $n$ items as input and jointly maps them to an $n$-dimensional vector of reals.
Each element in the output vector determines the \emph{relative} relevance of an item with respect to other items in the input.
While it is straightforward to define a multivariate function, it is less clear how such a function may be efficiently learned from training data or efficiently evaluated during inference given lists of arbitrary and variable number of items.
To that end, we propose Groupwise Scoring Function (GSF) as an instance of the class of multivariate functions that is parameterized by deep neural networks. A GSF learns to score a fixed-size "group" of items. We show how this model can be extended to act on lists of arbitrary length and demonstrate how efficient training and inference can be achieved by a Monte Carlo sampling strategy.
Empirical experiments on a private email search corpus and a public benchmark demonstrate that GSFs can achieve the state-of-the-art performance in learning-to-rank tasks. 

In particular, our contributions can be summarized as follows:
\vspace{-5pt}
\begin{itemize}
    \item We motivate and formulate multivariate scoring functions for learning-to-rank;
    \item We present Groupwise Scoring Function (GSF) as an instance of the class of multivariate functions that is parameterized by deep neural networks;
    \item We explore the conditions under which a GSF reduces to existing learning-to-rank models;
    \item We demonstrate, through empirical evaluation on proprietary and public datasets, the improvements obtained by GSFs and discuss their potential for learning-to-rank tasks;
    \item To encourage research in this space and to allow for reproducibility of the reported results, we open source our implementation within the TF Ranking library~\cite{TensorflowRanking2018}.
\end{itemize}


\section{Related Work}
\label{sec:related}

Learning-to-rank refers to algorithms that model the ranking problem with machine learning techniques. In general, ranking is formulated as a score-and-sort problem with the objective of constructing a scoring function where scores computed by such a function induce an ordering of items in a list. Existing learning-to-rank algorithms~\cite{joachims2006training, friedman2001greedy, burges2010ranknet, burges2005learning, cao2007learning,xia2008listwise} mainly differ by two factors: (a) the parameterization of the scoring function (e.g., linear functions~\cite{joachims2006training}, boosted weak learners~\cite{Jun+Hang:2007}, gradient-boosted trees~\cite{friedman2001greedy, burges2010ranknet}, support vector machines~\cite{joachims2006training, Joachims:WSDM17}, and neural networks~\cite{burges2005learning}); and (b) the loss function (e.g., pointwise~\cite{Gey:1994:IPR:188490.188560}, pairwise~\cite{burges2010ranknet,burges2005learning,joachims2006training} and listwise~\cite{cao2007learning,xia2008listwise}). Virtually all of the existing algorithms, however, yield a univariate scoring function in the end where the score of an item is computed in isolation and independently of other items in the list. To the best of our knowledge, there are only a few exceptions.

First, the score regularization technique~\cite{diaz2007regularizing} and the CRF-based model~\cite{qin2009global} use document similarities to smooth the initial ranking scores or enrich query-document pair feature vectors. When computing relevance scores, however, both methods take only one document at a time.

The second exception is a bivariate scoring function~\cite{Dehghani:SIGIR2017} that takes a pair of documents as input and predicts the preference of one over the other. It is easy to show that the bivariate scoring function is a special case of our proposed framework. 

Third is a group of neural learning to rank algorithms~\cite{ai2018learning,bello2018seq2slate} and click model~\cite{Borisov+al:2016} that builds an recurrent neural network over document lists. They, however, either focus on a re-ranking problem or use a pointwise loss to optimize user clicks.
In contrast, our method can be applied to arbitrary number of documents with any types of ranking loss functions.

Search result diversification is another area of related work. Diversification algorithms maximize objectives that take subsets of documents into account. These include maximal marginal relevance~\cite{Carbonell+Goldstein:1998} and subtopic relevance~\cite{Agrawal+al:2009}. Recently, several deep learning algorithms were proposed with losses corresponding to those objectives~\cite{Jiang+al:2017, Xia+al:2016}. 
In contrast, our work focuses on improving relevance, not diversity, by way of cross-document comparisons.

Another area of related research is the work on pseudo-relevance feedback~\cite{Manning:2008:IIR:1394399} where queries are expanded based on the top retrieved documents in a first round. The idea is that expanded queries lead to improvements in a second-stage retrieval. In this paper, we consider document relationships in the learning-to-rank setting, not retrieval, and do not require two rounds of retrieval. We also do not assume a pre-existing initial ordering of the document list.

Finally, note that our work is orthogonal and complementary to the recently proposed neural IR techniques ~\cite{DeepRank:2017,Guo+al:2016,Mitra+al:2017,Dehghani:SIGIR2017}. These techniques focus on advanced representations of document and query text but employ standard loss and scoring functions. On the other hand, our work concerns the nature of the scoring functions while employing a relatively simple query-doc representation.

\section{Problem Formulation}\label{sec:problem_formuation}

In this section, we formulate our problem in the context of learning-to-rank. Let $\psi=(\bm{x}, \bm{y}) \in \mathcal{X}^n \times \mathbb{R}^n$ be a training sample where $\bm{x}$ is a vector of $n$ items $x_i,\, 1\leq i \leq n$, $\bm{y}$ is a vector of real $n$ relevance labels $y_i,\, 1 \leq i \leq n$, and $\mathcal{X}$ is the space of all items. To simplify discussion and to follow convention, we refer to $x_i$ simply as a "document" and $\bm{x} \in \mathcal{X}^n$ as a list of $n$ documents, but note that $x_i$ itself could be a feature vector representing a query-document pair.
For every document $x_i \in \bm{x}$, we have a corresponding relevance label $y_{i} \in \bm{y}$. Finally, let $\Psi$ be a set of training examples. 

The goal of learning-to-rank can often be stated as finding a scoring function $f: \mathcal{X}^n \rightarrow \mathbb{R}^n$ that minimizes the empirical loss over the training data:
\begin{equation}
\mathcal{L}(f) = \frac{1}{|\Psi|}\sum_{(\bm{x}, \bm{y})\in\Psi} \ell(\bm{y},f(\bm{x})),
\label{equ:global_loss}
\end{equation}
where $\ell(.)$ is a local loss function.

As noted in earlier sections, the main difference between the various learning-to-rank algorithms lies in how the scoring function $f(\cdot)$ and the loss function $\ell(\cdot)$ are defined. While there are numerous examples of prior work on different types of loss functions~\cite{liu2009learning}, the vast majority of learning-to-rank algorithms assume a \emph{univariate} scoring function $u: \mathcal{X} \rightarrow \mathbb{R}$ that computes a score for each document independently of other documents:
\begin{equation}
f(\bm{x})|_i = u(x_i),\, 1 \leq i \leq n,
\end{equation}
where $f(\cdot)|_i$ denotes the $i^{\text{th}}$ dimension of $f$.

A score obtained from $u(\cdot)$ depends only on its argument. In other words, fixing $x_i$ and changing any or all other documents in the list to $x^{\prime}$ (for $j \neq i$) does not affect the output of $u(x_i)$.

In this paper, we set out to explore the space of \emph{multivariate} scoring functions $f: \mathcal{X}^n \rightarrow \mathbb{R}^n$ for learning-to-rank. Following our discussion in Section~\ref{sec:intro}, such a function is theoretically able to capture the relationship between its arguments and, as a result, could jointly produce \emph{relative} scores. In other words, replacing $x_i$ with a new document $x^{\prime}_i$ could lead to a change to scores for all documents in the list. Note, however, that any multivariate scoring function $f(\bm{x})$ should ideally be invariant to the order of items in $\bm{x}$.

Learning and evaluating a multivariate scoring function in practice is, however, nontrivial. In the discussion above, we made a simplifying assumption that $n$, the number of documents in a list, is constant across all training samples. As is common in learning-to-rank settings, however, that is often not the case and in fact the length of $\bm{x}$ is arbitrary and varies across training or evaluation samples. It is therefore not immediately clear how one may construct a generic multivariate function. In the following section, we address these challenges and introduce an instance of multivariate scoring functions that is suitable for the task of learning-to-rank and is further trained and evaluated in an efficient manner.


\section{Groupwise Scoring Functions} \label{sec:method}
In this section, we present a detailed construction of an instance of multivariate scoring functions which we refer to as \emph{groupwise scoring functions} (GSFs).
A GSF in its basic form is a function $g(\cdot; \theta): \mathcal{X}^m \rightarrow \mathbb{R}^m$ that is parameterized by a deep neural network (DNN) and that jointly maps a group of $m$ documents (where $m$ is fixed) to a vector of scores of the same size.
We begin this section by laying out the foundations of a GSF, later proceed to extend it to lists of $n \geq m$ documents, where $n$ may vary across samples, and finally complete the construction by providing a mechanism to efficiently train and evaluate an extended GSF.

\subsection{Parameterization by DNNs}\label{sec:dnn}

As noted earlier, we parameterize our functions using deep neural networks. Feed-forward neural networks have widely been applied to learning-to-rank problems~\cite{Edizel+al:2017, Huang+al:2013, Zamani+al:2017}. The reasons we believe a deep neural network fits well into our framework are two-fold. First, compared to tree models, neural networks scale well to high-dimensional inputs. This is important because a GSF takes $m$ documents as input where each document is a vector of an arbitrary and potentially large number of features. Second, neural networks arguably handle sparse features such as text more naturally whereas other models require extensive feature engineering. 
As such we believe a deep neural network is the right candidate for the task of learning a GSF.

To begin the construction, we need to define an input layer. Conceptually, a document $x$ can be represented as a concatenation of two subsets of features: the embedding features for sparse textual features $x^{\text{embed}}$ (e.g., for document titles) and the dense features $x^{\text{dense}}$ (e.g., document static scores or various match scores~\cite{Guo+al:2016}). For simplicity, we construct the input layer by concatenating all $m$ documents. Specifically, let 
$$\bm{h}_0 = \textrm{concat}(x_{1}^\text{embed}, x_{1}^\text{dense}, ..., x_{m}^\text{embed}, x_{m}^\text{dense}).$$
Note that in practice the input layer can be extended to include document-independent "context" features (such as query embeddings) and need not be limited to document-derived features.

Given the above input layer, we build a multi-layer feed-forward network with 3 hidden layers as follows:
\begin{equation}
\bm{h}_k = \sigma(\bm{w}_k^T\bm{h}_{k-1} + \bm{b}_k), ~~ k = 1,2,3
\label{equ:dnn}
\end{equation}
where $\bm{w}_k$ and $\bm{b}_k$ denote the weight matrix and the bias vector in the $k$-th layer, $\sigma$ is an activation function, which in this work is the ReLU function: $\sigma (t) = \text{max}(t, 0)$.

Our groupwise scoring function $g$ is thus defined as: 
\begin{equation}
g(\bm{x}) = \bm{w}_o^T\bm{h}_{3} + \bm{b}_o
\label{equ:pairwsie_output}
\end{equation}
where $\bm{w}_o$ and $\bm{b}_o$ are the weight vector and the bias in the output layer. The output layer of the network consists of $m$ units, each producing a score for each of the $m$ documents.

We note that in this work we wish to keep the design of our input layer and network architecture simple as these details, while important and consequential, are not germane to the topic of this work. We leave the exploration of more sophisticated representation of groups of input documents and advanced layers as future work.

\subsection{Extension to Arbitrarily Long Lists}\label{sec:group}

\begin{figure}
	\centering
	\includegraphics[width=3.4in]{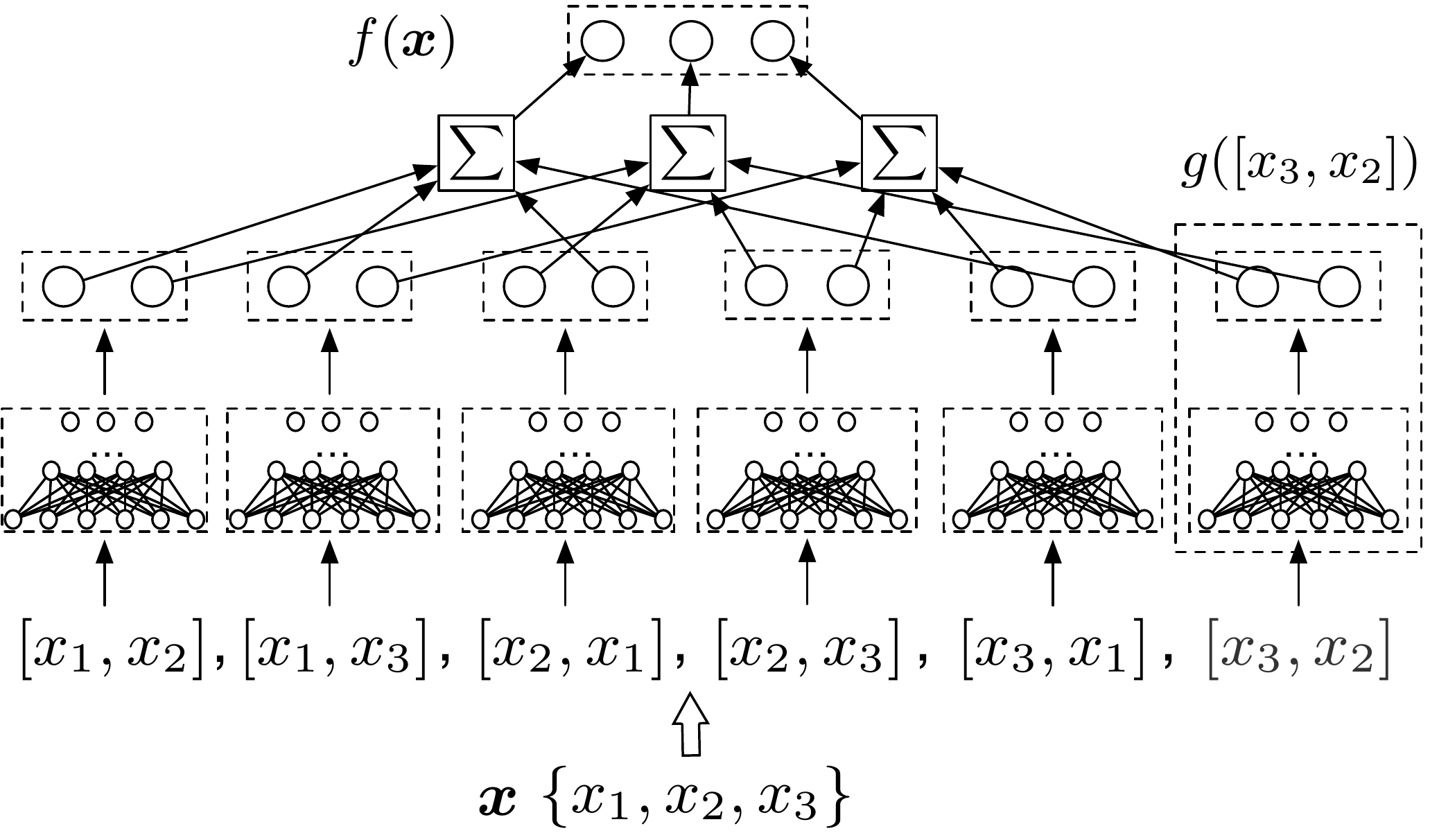}
	\vspace{-20pt}
	\caption{An extended groupwise scoring function. For illustrative purposes, we simplify $g(.)$ to be a bivariate function acting on permutations of size 2 formed from a list of 3 documents $\bm{x}$. All 6 size-2 permutations from $\bm{x}$ are fed to $g(.)$ which itself outputs 2 scores per permutation. Intermediate scores computed by $g$ are subsequently aggregated to compute the final vector of scores $f$.}
	\label{fig:groupwise}
\end{figure}

The domain of the function $g(\cdot)$ presented in the previous section is $\mathcal{X}^m$ with $m$ fixed. As noted earlier, in learning-to-rank, it is often the case that the list size (i.e., number of documents retrieved for a query) varies between queries. That important detail poses a challenge when designing and training a GSF.

Addressing this challenge by brute-force, one may set $m$ to be the corpus size and subsequently zero-pad input lists during training and inference. To state the obvious, the resulting network clearly does not scale to real-world corpora. Moreover, given the enormity of the parameter space, training such a network becomes prohibitive and any resulting model is unlikely to be effective.

A more viable solution, and one that we adopt to extend GSFs, is the following: Given a list of documents $\bm{x}$ of an arbitrary size $n$ and a GSF $g: \mathcal{X}^m \rightarrow \mathbb{R}^m$, we propose to compute $g(.)$ on size-$m$ permutations of $\bm{x}$ and accumulate scores along the way. 

Let $\Pi_m(\bm{x})$ denote a set of all possible $\frac{n!}{(n-m)!}$ permutations of size $m$ of the $n$ documents in $\bm{x}$, and let $\pi_k \in \Pi_m(\bm{x})$ be an element of that set. A permutation $\pi_k$ can be understood as a group of $m$ documents. In our proposed method, we compute $g(\cdot)$ on $\pi_k$ for all $k$. The vector of values $g(\pi_k)$ contains the scores of all documents $x_i \in \pi_k$ relative to other documents in that group.
Group scores $g$ are subsequently used to compute a final score for all $n$ documents. To explain that, it helps to define the following function:
\begin{equation}
    h(\pi, x) = \begin{cases}
    g(\pi)|_{\pi^{-1}(x)}, & \text{if } x \in \pi\\
    0, & \text{otherwise},
    \end{cases}
\end{equation}
where we use $\pi^{-1}(x)$ to denote the position of $x$ in $\pi$. The final score $f(\cdot)$ is then calculated by the following equation:

\begin{equation}
    f(\bm{x})|_i = \sum_{\pi_k \in \Pi_m(\bm{x})}{h(\pi_k, x_i)},\, 1 \leq i \leq n.
    \label{equ:final_score}
\end{equation}

Figure~\ref{fig:groupwise} illustrates one such $f(\cdot)$ in a simplified setting where $g(\cdot)$ is bivariate and $\bm{x}$ is a list of 3 documents.

\subsection{Efficient Training and Inference}\label{sec:inference}

One caveat of the extended GSF is the factorial growth of the space of permutations $\Pi_m(.)$. For large values of $n$, the set $\Pi_m(\bm{x})$ grows so intractably large that computing $g(.)$ on the resulting groups and aggregating group scores by Equation~\ref{equ:final_score} quickly become prohibitive: assuming the computational complexity of $g(\cdot)$ is $\mathcal{O}(m)$ such a scoring paradigm has a complexity of $\mathcal{O}(m \frac{n!}{(n - m)!})$.

To reduce the complexity of GSFs, we propose to substitute the summation in Equation~\ref{equ:final_score} with an expectation as follows:
\begin{equation}
    f(\bm{x})|_i = \mathbb{E}_{\pi \ni x_i}[g(\pi, x_i)|_{\pi^{-1}(x_i)}],\, 1 \leq i \leq n.
    \label{equ:expected_final_score}
\end{equation}

The expectation in Equation~(\ref{equ:expected_final_score}) can be approximated effectively using Monte Carlo methods~\cite{MonteCarlo}. In our implementation, we use the following sampling recipe: From each training sample with document list $\bm{x}$, we form groups by taking sub-sequences of a randomly shuffled version of $\bm{x}$.

Such down-sampling substantially reduces the time complexity to $\mathcal{O}(mn)$. It is easy to show that, because each $x_i \in \bm{x}$ appears in exactly $m$ groups, each document is equally likely to be compared with other documents in the list. Moreover, a document's position in the group is also uniformly distributed. Given enough training data, a GSF trained using this sampling strategy asymptotically approaches a GSF trained with all permutations and is further invariant to document order in the input list.

\subsection{Loss Function}

We train a GSF by optimizing the empirical loss in Equation~(\ref{equ:global_loss}) using back-propagation. While in theory any arbitrary loss function $\ell(\cdot)$ can be used within this framework---more on this in Section~\ref{sec:relationship_with_others}---we empirically found the cross-entropy loss to be particularly effective. We define this loss as follows:
\begin{equation}
\ell(\bm{y},f(\bm{x})) = - \sum_{i=1}^{n} \frac{y_i}{Y} \cdot \log p_i
\label{equ:softmax_loss}
\end{equation}
where $Y = \sum_{y\in\bm{y}}{y}$ is a normalizing factor, and $p_i$'s are the projection of scores $f(\bm{x})$ onto the probability simplex using Softmax:
\begin{equation}
\mathit{Softmax}(\bm{t})|_i = \frac{e^{t_i}}{\sum_{j=1}^{n} e^{t_j}},\, 1 \leq i \leq n.
\label{equ:softmax}
\end{equation}

An important property of this loss function is that it can be incorporated into an unbiased learning-to-rank framework. Specifically, it is easy to extend this loss to factor in Inverse Propensity Weights~\cite{Joachims:WSDM17, Wang+al:2018} to counter position bias in click logs. The IPW-enabled variant of the loss in Equation~(\ref{equ:softmax_loss}) is as follows:
\begin{equation}
\ell(\bm{y},f(\bm{x})) = - \sum_{i=1}^{n} w_i \cdot y_i \cdot \log p_i = - \sum_{i: y_i = 1} w_i \cdot \log p_i,
\label{equ:ipw_softmax_loss}
\end{equation}
where $w_i$ is the Inverse Propensity Weight of the $i^{\text{th}}$ result, and where it is assumed that $y \in \{0,\, 1\}$ for click logs and that only one document is clicked (i.e., $Y=1$).

We use the above loss in the experiments reported in this work and leave the exploration of more advanced loss functions as future work. We will also defer a theoretical analysis of the cross-entropy loss or its extensions in the context of GSFs to a future study.


\subsection{Relationship with Existing Models}\label{sec:relationship_with_others}
In this section, we discuss the relationship between some of the existing learning-to-rank algorithms and our proposed model. In particular, a GSF model can be reduced to most existing algorithms by way of tuning a few knobs including group size $m$, loss function $\ell(\cdot)$, and the score aggregation function $f(\cdot)$. This includes RankNet~\cite{burges2010ranknet}, ListNet~\cite{cao2007learning}, and the work by Dehghani et al.~\cite{Dehghani:SIGIR2017}. Due to space limitation, we only show the ListNet as an example.

A traditional listwise model uses a univariate scoring function with a listwise loss that is computed over all documents in the list. 
It is easy to see how a GSF can be modified and reduced to a univariate function with a listwise loss: Fix $m=1$ for $g(\cdot)$, define $f(\cdot)$ as in Equation~(\ref{equ:expected_final_score}), and plug any listwise loss $\ell(\cdot)$ into Equation~(\ref{equ:global_loss}).

Let us go through this exercise by presenting a configuration that transforms our GSF model to ListNet~\cite{cao2007learning}. Given the univariate nature of $g(\cdot)$ in the new configuration, define $\hat{\bm{y}}$ as 
\begin{equation}
    \hat{\bm{y}} \triangleq f(\bm{x}),~~~~~~~\hat{y}_i \triangleq f(\bm{x})|_i = g(x_i),\, 1 \leq i \leq n.
    \label{equ:univariate_y_hat}
\end{equation}
ListNet optimizes the cross-entropy loss between two ("top-one" probability) distributions: One obtained from relevance labels $\bm{y}$ and another defined over scores $\hat{\bm{y}}$. The following expression defines the ListNet loss:
\begin{equation}
\ell(\bm{y}, \hat{\bm{y}}) = - \sum_{i=1}^{n}\frac{e^{y_i}}{\sum_{j=1}^{n}{e^{y_j}}}\log \frac{e^{\hat{y}_i}} {\sum_{j=1}^{n} e^{\hat{y}_j}}.
\end{equation}
Using the above loss in Equation~(\ref{equ:global_loss}) completes the transformation of a GSF to the ListNet model.

Note that the ListNet loss is almost identical to the loss used in our GSF model as shown in Equation~(\ref{equ:softmax_loss}). ListNet, however, projects labels to the probability simplex using the Softmax function whereas in GSF labels are simply normalized. When $y_i = 0$, we calculate zero loss in the GSF setup while this is not the case in the standard ListNet loss; the ListNet loss is always non-zero. This difference becomes important if one wishes to train a model in an unbiased learning-to-rank framework~\cite{Wang+al:2016,Joachims:WSDM17} where propensity weights cannot be computed for non-clicked documents~\cite{Wang+al:2018}. As such, having a non-zero loss for non-clicked documents proves to be a significant limitation of the ListNet loss in the context of unbiased learning.

\section{Experimental Setup}\label{sec:eval}

\begin{table}\small
	\centering
	\caption{List of baseline DNN models.}
	\scalebox{0.9}{
	\begin{tabular}{p{1.5cm} | p{6.2cm} } %
		\hline
		\textsc{PointDNN} & A standard DNN model with a univariate scoring function and pointwise loss~\cite{Zamani+al:2017}. \\ \hline
		RankNet & A neural network model with univariate scoring and pairwise loss~\cite{burges2005learning}. \\ \hline
        \textsc{BiDNN} & The standard DNN model with bivariate scoring and Sigmoid cross entropy~\cite{Dehghani:SIGIR2017}. \\
		\hline 
	\end{tabular}
	}
	\label{tab:models}
\end{table}

\begin{table}[t] \small
	\vspace{-22pt}
	\centering
	\caption{List of GSF variants.}
	\scalebox{0.9}{
	\begin{tabular}{p{1.5cm} | p{6.2cm} } %
		\hline
		\textsc{PairGSF} & GSF reduced to a univariate scoring function with a pairwise loss used in RankNet~\cite{burges2010ranknet}. \\ \hline
        \textsc{BiGSF} & GSF reduced to a bivariate scoring function similar to~\cite{Dehghani:SIGIR2017},
        but where the aggregation function remains as in Equation~\ref{equ:expected_final_score}. \\ \hline
        \textsc{GSF}($m$) & GSF model with group size $m$.  \\
		\hline
	\end{tabular}
	}
	\label{tab:gsf-models}
\end{table}

GSFs have theoretically interesting properties but their effectiveness in practice remains to be verified empirically. In the remainder of this paper, we set out to do just that by evaluating our proposed method on two datasets. To conduct experiments, we have implemented the GSF model in Tensorflow, a standard deep learning platform. In order to facilitate reproducibility of the reported results, we open source our code within the the TF Ranking library~\cite{TensorflowRanking2018}. Moreover, in this section, we give a detailed description of our experimental design, setup, and model hyper-parameters.

\subsection{Baseline Learning-to-Rank Models}

We compare our method with a number of existing learning-to-rank algorithms that fall into two categories: DNN models and tree-based models.
Table~\ref{tab:models} summarizes a list of DNN models we use as baselines in our experiments. In this table, \textsc{PointDNN} and RankNet represent the existing DNN models with a univariate scoring function in the learning-to-rank literature. \textsc{BiDNN} is a recently proposed model that takes a pair of documents and jointly computes preference scores~\cite{Dehghani:SIGIR2017}.
As for tree-based models, we primarily use the state-of-the-art MART and LambdaMART~\cite{burges2010ranknet} algorithms as a baseline to compare with. In general, tree-based models cannot efficiently handle high-dimensional sparse features such as document text. Therefore, where we compare DNNs and tree-based models we do so by training a model with dense features only. Furthermore, where possible, we also explore a hybrid approach in which predictions from the DNN models are used as input features for tree-based models. Such a hybrid approach enables us to incorporate sparse features into tree-based models; we compare hybrid models with both standalone DNN and tree-based models in our experiments.

For completeness, Table~\ref{tab:gsf-models} summarizes the different GSF variants considered in the following sections.

\subsection{Datasets} \label{sec:data}

We conduct a first set of experiments on a click dataset that is obtained from search logs of one of the largest commercial email search engines. In this service, a maximum of 6 results are returned and presented to users in an overlay. The overlay disappears after a click and the clicked result is then displayed. As a result, at most one click is obtained per query session. For this dataset, we discard all sessions that do not contain a click. For sessions with a click, we keep all 6 displayed documents and their click/no-click is recorded as relevance labels. This process results in approximately 150 million sessions in total. We sampled 5 million sessions to construct a held-out test set and used the rest for training and validation with a 9 : 1 ratio. To train \textsc{BiDNN} and \textsc{BiGSF}, we sample all pairs where one document is clicked.

The features in this dataset consist of both dense and sparse features. The dense features include query-document matching features like BM25. These types of features are the primary features used in traditional learning-to-rank algorithms~\cite{liu2009learning}. Recently, sparse features were shown to be effective through embedding in an end-to-end deep neural network model~\cite{Dehghani:SIGIR2017}. Our click dataset contains n-grams from query strings and document subjects as sparse features. The average of the embedding vectors for n-grams in a query or document subject is used as the feature representation. 

The second dataset used in our experiments is the publicly available MSLR-WEB30K~\cite{DBLP:journals/corr/QinL13}. This is a large-scale learning-to-rank set that contains 30,000 queries. On average there are 120 documents per query and each document has 136 numeric features. All documents are labeled with graded relevance from 0 to 4 with larger labels indicating a higher relevance. We evaluate the models on Fold~1 of this dataset. Results obtained on the other folds are similar.

\subsection{Hyperparameters and Training}
We build the DNN models using a 3-layer feed-forward network. On the Email dataset, hidden layer sizes are set as 256, 128 and 64 for $\bm{h}_1$, $\bm{h}_2$ and $\bm{h}_3$ respectively. We set the learning rate to 0.1 and training batch size to 100. For sparse features, we set the embedding dimension to 20. Larger embedding dimensions (e.g.,~100), learning rates (e.g., 0.2, 0.3), and layer sizes (e.g.,~512 or 1024) were tested but no significant difference was observed. For this dataset, we use unbiased learning-to-rank techniques to overcome click bias~\cite{Joachims:WSDM17, Wang+al:2018} and to that end, we optimize the weighted variant of the cross-entropy loss as shown in Equation~(\ref{equ:ipw_softmax_loss}) during training.

In models trained on Web30K, hidden layers have 64, 32, and 16 units instead with batch normalization between consecutive layers. A learning rate of 0.005 is used and training batch size is set to 128. These hyper-parameters were found to be effective through fine-tuning on the validation set. We train the model for 30,000 steps and evaluate the final model. Finally, when aggregating losses from queries in a mini-batch, a query's loss is weighted by the sum of the relevance grade of its documents.

For both datasets, we use Adagrad to optimize the objective.

\subsection{Evaluation}\label{sec:unbiased_evaluation}
For experiments on the Email dataset, we report an Inverse Propensity Weight (IPW) enabled variant of mean reciprocal rank (MRR)~\cite{Wang+al:2016}. Such a weighted metric allows us to correct for the position bias that exists in click logs. Let $N$ denote the number of test sessions and $\mathit{rank}_i$ be the rank of the clicked document for the $i^{\text{th}}$ session, then weighted MRR is calculated as follows:
\begin{equation}
\mathit{WMRR} = \frac{1}{\sum_{i=1}^{N} w_i}\sum_{i = 1}^N\frac{w_i}{\mathit{rank}_i}
\label{equ:WMRR}
\end{equation}
where $w_i$ is the IPW of the clicked document for the $i^{\text{th}}$ session.

For experiments on the Web30K dataset, we run 10 trials of every model configuration and report mean Normalized Discounted Cumulative Gain~\cite{jarvelin2002cumulated} at rank positions 1, 5, and 10 along with $95\%$ confidence intervals. Note that when computing NDCG, queries with no relevant documents are discarded from the evaluation set. 
Also, each trial may produce a different model given the same hyper-parameters due to the stochastic nature of network initialization, as well as batch-level and query-level shuffling of documents.

\section{Experimental Results}\label{sec:results}

In this section, we report the results of our experiments. We first examine the effect of group sizes $m$ on GSF models.
We then compare GSFs with the state-of-the-art learning-to-rank algorithms.

\begin{figure}[t]
	\centering
	\includegraphics[width=2.4in]{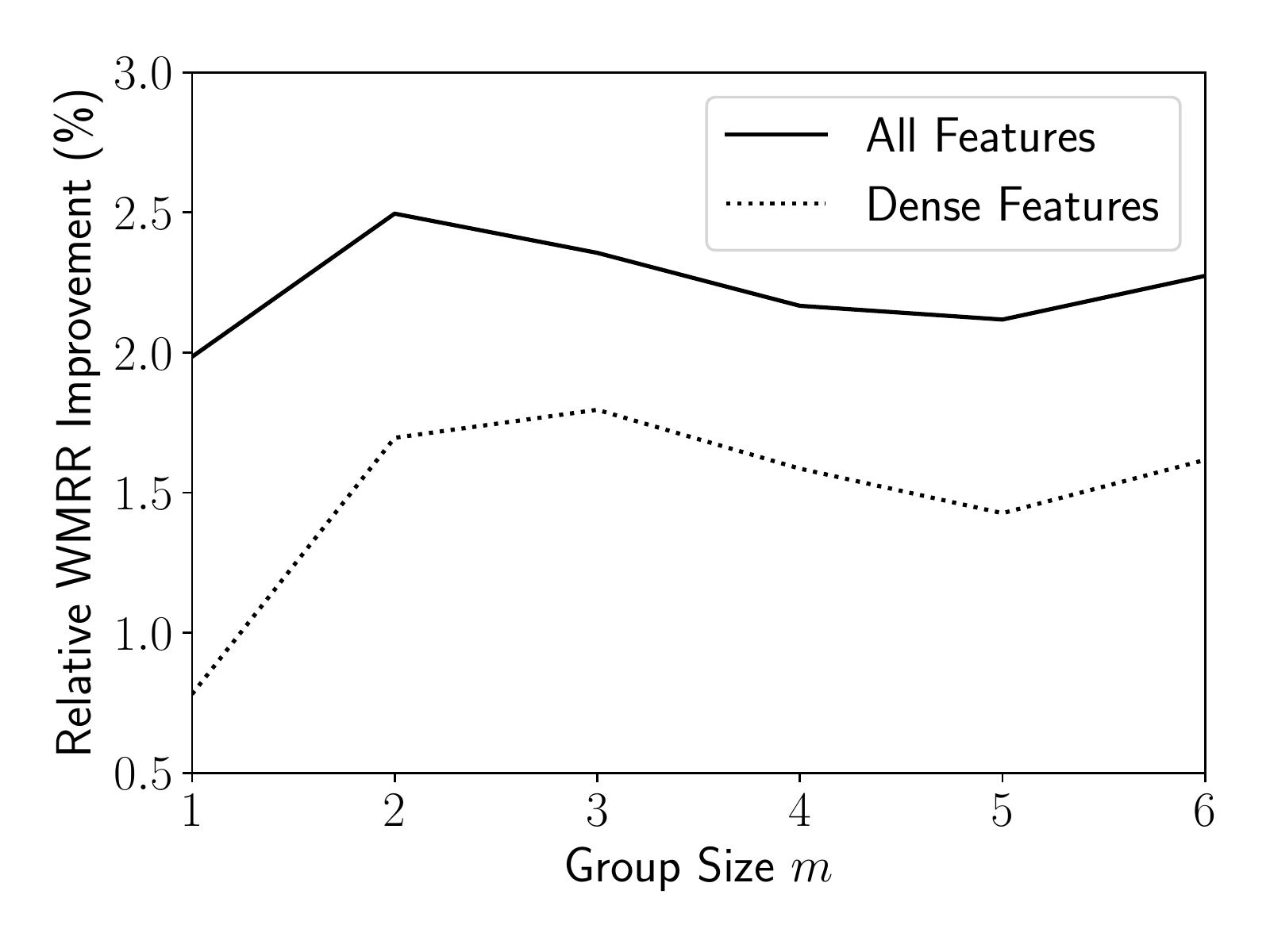}
	\vspace{-15pt}
	\caption{Relative WMRR improvement over \textsc{PointDNN}~for GSF models with different group sizes using dense features and all (dense and sparse) features on the Email dataset.}
	\label{fig:group-size}
\end{figure}

\begin{figure}[t]
	\centering
	\includegraphics[width=2.4in]{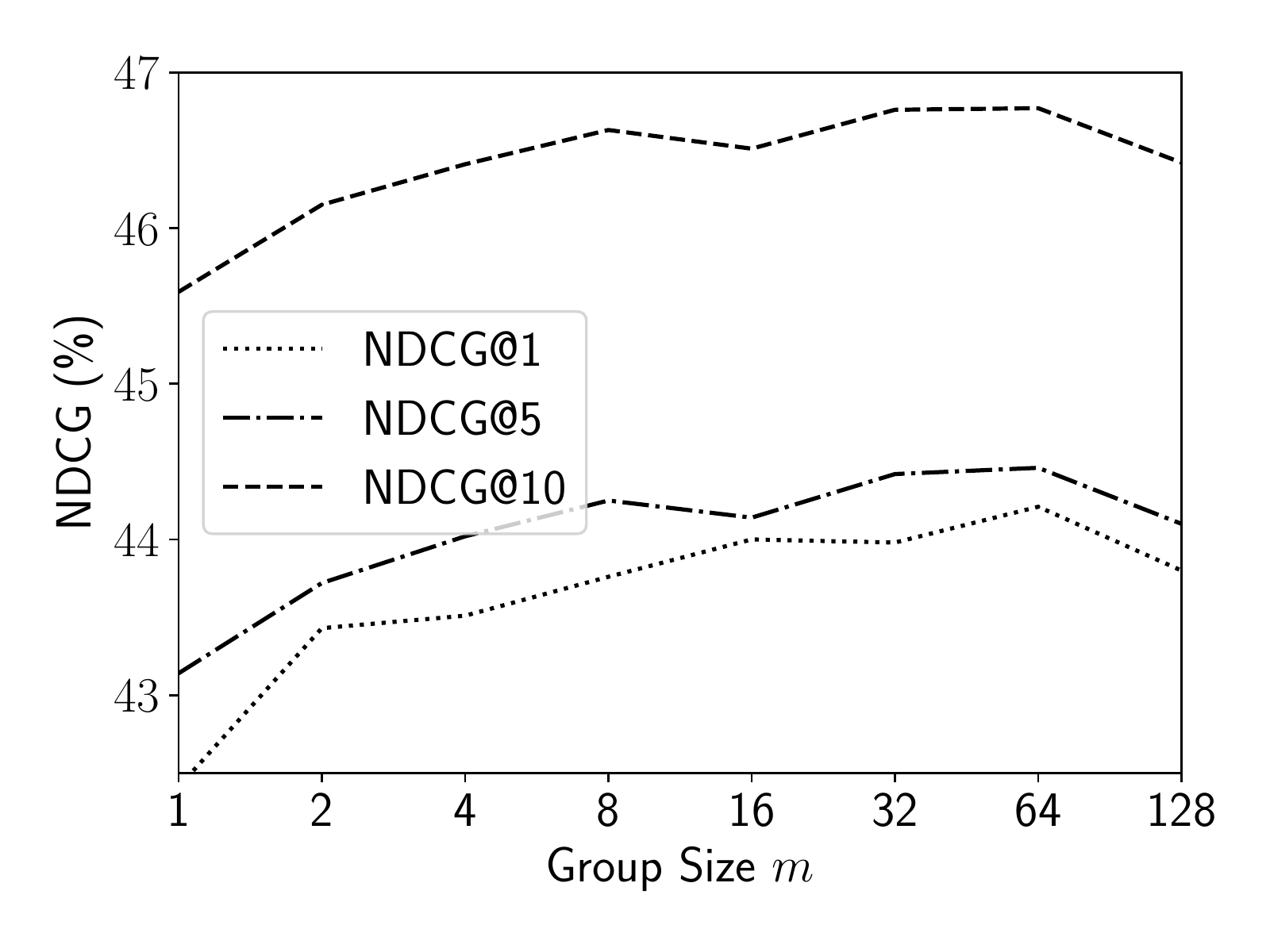}
	\vspace{-15pt}
	\caption{NDCG at rank positions 1, 5, and 10 (in percentage) for GSFs with different group sizes on the Web30K dataset.}
	\label{fig:web30K-group-size}
\end{figure}

\subsection{Effect of Group Size}

As discussed in Sections~\ref{sec:group} and \ref{sec:inference}, while a GSF can easily and efficiently extend to a variable list size $n$, the group size $m$ must be fixed before the construction of the model. 
To study the effect $m$ has on resultant models, we conduct experiments with different configurations on both the Email and Web30K datasets. 

On the Email dataset, we train a \textsc{PointDNN}\ model (a univariate scoring function with pointwise loss, see Table~\ref{tab:models}) as baseline. We then measure improvements over this model, as indicated by WMRR, of GSF models trained with different group sizes. We repeat these experiments for two settings: (a) \emph{all features} experiments use both sparse query and document textual features as well as numerical features; and, (b) \emph{dense features} experiments use only the dense numerical features. Results are illustrated in Figure~\ref{fig:group-size}.

From Figure~\ref{fig:group-size}, we can see that a GSF trained with all features reaches its peak performance when $m=2$, and GSF with dense features reaches its peak when $m=3$. We also observe that the ranking quality decreases slightly when the group size becomes larger. 
We believe this observation can be explained by the fact that feed-forward networks usually are sensitive to the input order. 
As group size increases, the number of permutations for a group of documents grows rapidly. 
When this happens, because of the particular sampling process we use to form groups from a document list (see Section~\ref{sec:inference} for details), the approximation of the expectation in Equation~(\ref{equ:expected_final_score}) becomes less accurate.

On the Web30K dataset, we measure the performance of GSF models in terms of NDCG at rank positions 1, 5, and 10 and report these metrics for various group sizes. Figure~\ref{fig:web30K-group-size} illustrates the results for groups of size 1, 2, 4, 8, 16, 32, 64, and 128. We observe an upward trend as we increase the group size $m$. The models with group size $m \geq 8$ are approximately 3\% better than those with group size 1. Noting that there are only dense features in the Web30K dataset, this indicates that the extension of univariate scoring functions to multivariate scoring functions could be particularly useful for learning-to-rank models with dense features. Once again, for models with very large group sizes ($m \geq 32$), we observe that NDCG plateaus or drops slightly, which can be explained by inadequate sampling and the growth of the space of permutations.

\subsection{Comparison with Baseline DNNs}

\begin{figure}[t]
	\centering
	\includegraphics[width=2.4in]{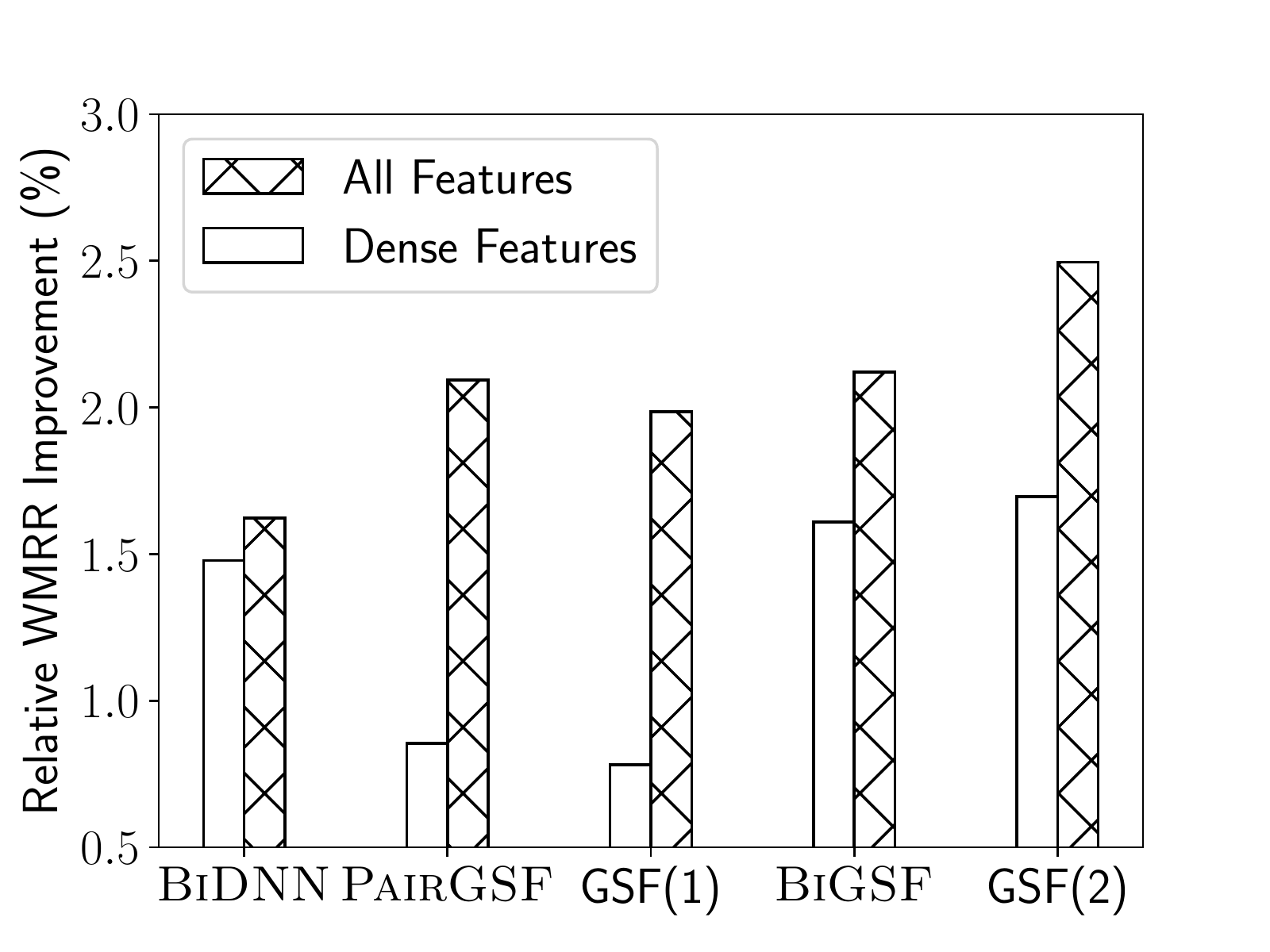}
	\caption{Relative WMRR improvements over \textsc{PointDNN}~on all features and dense features on the Email dataset. All improvements over \textsc{PointDNN}~are statistically significant according to a \emph{t}-test with $\alpha < 0.01$. The improvement of GSF with group size 2, denoted GSF(2), over the other models is also statistically significant at $\alpha < 0.05$.}
	\label{fig:results-on-click-data}
\end{figure}

We are interested in the relative performance of GSF models when compared with other baseline DNN algorithms. On the Email dataset, we measure the gain in WMRR over the \textsc{PointDNN}~method obtained by \textsc{PairGSF} (univariate GSF with pairwise loss),
\textsc{BiGSF} (bivariate GSF),
and finally GSF models with group sizes 1 and 2---we denote the last two models as GSF(1) and GSF(2) for brevity. To provide a reference point for how well GSF models perform, we also report the gain from the \textsc{BiDNN} model. The results of these comparisons on Email data is illustrated in Figure~\ref{fig:results-on-click-data}.

All five models achieve significant improvements over the \textsc{PointDNN}~baseline. 
Among them, GSF(2) yields the highest WMRR for both "all" and "dense" feature settings.
For example, using all features GSF(2) achieves a 2.5\% improvement over the \textsc{PointDNN}~baseline. This improvement is significantly better than the gain from \textsc{BiDNN}, an improvement of less than 1.5\%.

If we consider dense features only, \textsc{BiDNN}, \textsc{BiGSF}, and GSF(2) models---all bivariate functions---lead to better results than GSF(1), a univariate function.
This suggests that scoring documents jointly proves particularly effective when only dense features are available. 


We also compared \textsc{PairGSF}, \textsc{BiGSF}, GSF(1) and GSF(2) in terms of NDCG@5 on the Web30k dataset. Results are shown in Table~\ref{tab:dnn-vs-tree-on-letor}(a) which confirm again that the GSF is indeed more effective than the univariate and bivariate scoring functions.

\subsection{Comparison with Tree-based Models}

We next compare the proposed GSF models with tree-based models in both a standalone and a hybrid approach. In the hybrid setting---henceforth, referred to as LambdaMART+GSF---the output of the GSF model is used as a feature in LambdaMART.
We use LambdaMART as reference because it has been shown to yield state-of-the-art performance in public learning-to-rank competitions \cite{Chapelle+Chang:2011}.

Table~\ref{tab:dnn-vs-tree-on-click} shows the results we obtained on the Email dataset. For scalability reasons, we use an internal implementation of LambdaMART on this dataset.  As LambdaMART cannot natively handle raw textual features, we only report the relative improvement over LambdaMART with dense features. Based on the results in Figure~\ref{fig:results-on-click-data}, we use GSF(2) as the representative GSF model for this experiment. 

From Table~\ref{tab:dnn-vs-tree-on-click}, we see that GSF significantly outperforms LambdaMART in (a) dense features regime (where GSF slightly outperforms LambdaMART), and (b) all features regime (where the performance gap is much more significant). This demonstrates the importance of incorporating raw textual features and the effectiveness of GSF models in leveraging them. Furthermore, the hybrid LambdaMART+GSF approach achieves an even better performance, reaching gains as large as 3.42\% over LambdaMART, a statistically significant improvement over all other models in Table~\ref{tab:dnn-vs-tree-on-click}. 
This validates the complementary nature of our method to LambdaMART and the benefits of the hybrid approach.

\begin{table}[t] \small
	\centering
	\caption{Relative WMRR improvement over LambdaMART on the Email dataset. $*$ and $+$ denotes statistically significant improvements over LambdaMART with dense features and all other models in the table, both using \emph{t}-test with $\alpha < 0.01$.}
	\label{tab:dnn-vs-tree-on-click}
	\scalebox{0.9}{
	\begin{tabular}{ | p{4cm} | c | c|} 
		\hline & Dense Features & All Features \\
		\hline LambdaMART & 0.00\% & --\\
		\hline GSF(2) & 0.30\%$^*$ & 2.40\%$^*$ \\
		\hline LambdaMART+GSF(2) & 0.95\%$^*$ & \textbf{3.42\%}$^{*+}$ \\
		\hline
	\end{tabular}
	}
\end{table}

Table~\ref{tab:dnn-vs-tree-on-letor} shows the results on the Web30K dataset.  For reproducibility, we use several learning-to-rank models implemented in the open-source Ranklib toolkit\footnote{https://sourceforge.net/p/lemur/wiki/RankLib/} as baselines.

We observe that all GSF variants in Table~\ref{tab:dnn-vs-tree-on-letor}(a), outperform RankNet and RankSVM by a very large margin. A comparison of GSFs with MART and LambdaMART is shown in Table~\ref{tab:dnn-vs-tree-on-letor}(b), where we report mean NDCG at various rank positions over 10 trials along with 95\% confidence intervals. From the table, it is clear that the GSF setting with group size $m=64$ yields statistically significant improvements over MART at all NDCG cut-offs. On the other hand, GSF(64) falls short of LambdaMART as measured by NDCG@1, is on par in terms of NDCG@5 (i.e., confidence intervals overlap), and performs significantly better than LambdaMART as indicated by NDCG@10.
For completeness, we select the trial with the highest NDCG@1 on the validation set and measure its NDCG@1 on the test set. We repeat this for NDCG@5 and NDCG@10 and report the results in Table~\ref{tab:dnn-vs-tree-on-letor}(c). The conclusions from Table~\ref{tab:dnn-vs-tree-on-letor}(b) still hold.

The results from Table~\ref{tab:dnn-vs-tree-on-letor} are interesting. We believe the reason LambdaMART performs better than GSF at NDCG@1 is a result of the differences between loss functions: in the existing GSF setup, we use the cross-entropy loss which is not position-dependent, whereas LambdaMART's loss is designed to take position into account.

Our observations from a comparison of GSFs with tree-based models lead us to believe that the ranking quality of GSFs is at least on par with state-of-the-art tree-based models. As the number of training examples increases by orders of magnitude and when sparse textual features are present in a feature set, GSFs yield higher quality models and prove more scalable.

\begin{table}[t] \small
	\centering
	\caption{A comparison on the Web30K dataset of (a) various GSF flavors and weaker baselines by NDCG@5; (b) strong baseline models and the best-performing GSF variant by NDCG at different cut-offs with 95\% confidence intervals from 10 trials; and, (c) highest performing trial as measured by NDCG at different rank positions on the validation set. $^*$ denotes statistically significant differences between GSF and LambdaMART using \emph{t}-test with $\alpha < 0.05$.}
	\vspace{-5pt}
	\label{tab:dnn-vs-tree-on-letor}
	\scalebox{0.9}{
	\begin{tabular}{ | c | c | c | c | c | c | c |} 
		\multicolumn{7}{c}{\textbf{(a)}} \\
		\hline RankNet & RankSVM & \textsc{PairGSF} & \textsc{BiGSF} &  GSF(1) & GSF(2) & GSF(64)\\
		\hline 32.28 & 34.79 & 40.40 & 41.10 & 43.14 & 43.72 & \textbf{44.46} \\
		\hline
    \end{tabular}
    }
    \scalebox{0.9}{
    \begin{tabular}{ |l | c | c | c |}
		\multicolumn{4}{c}{\textbf{(b)}} \\
		\hline & MART & LambdaMART & GSF(64) \\
		\hline NDCG@1 & 43.73 ($\pm$0.01) & \textbf{45.35} ($\pm$0.06) & 44.21 ($\pm$0.18) \\
		\hline NDCG@5 & 43.96 ($\pm$0.03) & \textbf{44.59} ($\pm$0.04) & 44.46 ($\pm$0.12) \\
		\hline NDCG@10 & 46.40 ($\pm$0.02) & 46.46 ($\pm$0.03) & \textbf{46.77} ($\pm$0.13) \\
        \hline
		\multicolumn{4}{c}{\textbf{(c)}} \\
		\hline & MART & LambdaMART & GSF(64) \\
		\hline NDCG@1 & 43.76 & \textbf{45.27}${^*}$ & 44.47 \\
		\hline NDCG@5 & 44.03 & 44.56 & \textbf{44.63} \\
		\hline NDCG@10 & 46.44 & 46.52 & \textbf{47.01}$^{*}$ \\
		\hline
	\end{tabular}
	}
\end{table}

\section{Discussion and Future Work}\label{sec:discussion}
We began this work by stating a hypothesis, that the \emph{relative} relevance of an item would be more accurately estimated if relevance scores for all items were computed jointly. Experiments conducted in the last section shed light on that hypothesis and the questions raised earlier in this work. The results are encouraging.



GSFs, while not yet a fully mature deep learning framework, provide a blueprint for designing multivariate scoring functions for ranking. Analogous to the use of Recurrent Neural Networks in Natural Language Processing and Convolutional Neural Networks in Computer Vision, we believe GSFs are inspired by and are more appropriate for ranking, where relativity plays a large role. Moreover, GSFs incorporate local feature distributions by simultaneously considering multiple candidate documents, mimicking user behavior more closely. Thus, we believe that GSFs provide an opportunity for the advancement of learning-to-rank research using deep learning.

There are, for example, many components that warrant a closer look. A na\"ive concatenation of a list of input documents, as done in this work, may not be effective at preserving documents' structure and may lead to a loss of signals useful for comparison of documents. A feed-forward network may not be appropriate for capturing similarities or differences between documents. Finally, while the cross-entropy loss proved effective in practice, (a) it lacks theoretical justification and (b) a metric-driven loss may lead to better overall performance. We plan to pursue this direction of research and continue to improve our understanding of multivariate scoring and ranking functions. In order to facilitate and share this research, we open source GSFs within the TF Ranking library~\cite{TensorflowRanking2018}.




\section{Acknowledgments}
This work would not be possible without the support provided by the TF-Ranking team.

\balance
\bibliographystyle{ACM-Reference-Format}
\bibliography{paper}


\begin{thebibliography}{39}


\ifx \showCODEN    \undefined \def \showCODEN     #1{\unskip}     \fi
\ifx \showDOI      \undefined \def \showDOI       #1{#1}\fi
\ifx \showISBNx    \undefined \def \showISBNx     #1{\unskip}     \fi
\ifx \showISBNxiii \undefined \def \showISBNxiii  #1{\unskip}     \fi
\ifx \showISSN     \undefined \def \showISSN      #1{\unskip}     \fi
\ifx \showLCCN     \undefined \def \showLCCN      #1{\unskip}     \fi
\ifx \shownote     \undefined \def \shownote      #1{#1}          \fi
\ifx \showarticletitle \undefined \def \showarticletitle #1{#1}   \fi
\ifx \showURL      \undefined \def \showURL       {\relax}        \fi
\providecommand\bibfield[2]{#2}
\providecommand\bibinfo[2]{#2}
\providecommand\natexlab[1]{#1}
\providecommand\showeprint[2][]{arXiv:#2}

\bibitem[\protect\citeauthoryear{Agrawal, Gollapudi, Halverson, and
  Ieong}{Agrawal et~al\mbox{.}}{2009}]%
        {Agrawal+al:2009}
\bibfield{author}{\bibinfo{person}{Rakesh Agrawal}, \bibinfo{person}{Sreenivas
  Gollapudi}, \bibinfo{person}{Alan Halverson}, {and} \bibinfo{person}{Samuel
  Ieong}.} \bibinfo{year}{2009}\natexlab{}.
\newblock \showarticletitle{Diversifying Search Results}. In
  \bibinfo{booktitle}{\emph{Proc. of the 2nd ACM International Conference on
  Web Search and Data Mining}}. \bibinfo{pages}{5--14}.
\newblock


\bibitem[\protect\citeauthoryear{Ai, Bi, Guo, and Croft}{Ai
  et~al\mbox{.}}{2018}]%
        {ai2018learning}
\bibfield{author}{\bibinfo{person}{Qingyao Ai}, \bibinfo{person}{Keping Bi},
  \bibinfo{person}{Jiafeng Guo}, {and} \bibinfo{person}{W~Bruce Croft}.}
  \bibinfo{year}{2018}\natexlab{}.
\newblock \showarticletitle{Learning a deep listwise context model for ranking
  refinement}. In \bibinfo{booktitle}{\emph{The 41st International ACM SIGIR
  Conference on Research \& Development in Information Retrieval}}. ACM,
  \bibinfo{pages}{135--144}.
\newblock


\bibitem[\protect\citeauthoryear{Bello, Kulkarni, Jain, Boutilier, Chi, Eban,
  Luo, Mackey, and Meshi}{Bello et~al\mbox{.}}{2018}]%
        {bello2018seq2slate}
\bibfield{author}{\bibinfo{person}{Irwan Bello}, \bibinfo{person}{Sayali
  Kulkarni}, \bibinfo{person}{Sagar Jain}, \bibinfo{person}{Craig Boutilier},
  \bibinfo{person}{Ed Chi}, \bibinfo{person}{Elad Eban},
  \bibinfo{person}{Xiyang Luo}, \bibinfo{person}{Alan Mackey}, {and}
  \bibinfo{person}{Ofer Meshi}.} \bibinfo{year}{2018}\natexlab{}.
\newblock \showarticletitle{Seq2slate: Re-ranking and slate optimization with
  rnns}.
\newblock \bibinfo{journal}{\emph{arXiv preprint arXiv:1810.02019}}
  (\bibinfo{year}{2018}).
\newblock


\bibitem[\protect\citeauthoryear{Borisov, Markov, de~Rijke, and
  Serdyukov}{Borisov et~al\mbox{.}}{2016}]%
        {Borisov+al:2016}
\bibfield{author}{\bibinfo{person}{Alexey Borisov}, \bibinfo{person}{Ilya
  Markov}, \bibinfo{person}{Maarten de Rijke}, {and} \bibinfo{person}{Pavel
  Serdyukov}.} \bibinfo{year}{2016}\natexlab{}.
\newblock \showarticletitle{A Neural Click Model for Web Search}. In
  \bibinfo{booktitle}{\emph{Proc. of the 25th International Conference on World
  Wide Web}}. \bibinfo{pages}{531--541}.
\newblock


\bibitem[\protect\citeauthoryear{Burges, Shaked, Renshaw, Lazier, Deeds,
  Hamilton, and Hullender}{Burges et~al\mbox{.}}{2005}]%
        {burges2005learning}
\bibfield{author}{\bibinfo{person}{Chris Burges}, \bibinfo{person}{Tal Shaked},
  \bibinfo{person}{Erin Renshaw}, \bibinfo{person}{Ari Lazier},
  \bibinfo{person}{Matt Deeds}, \bibinfo{person}{Nicole Hamilton}, {and}
  \bibinfo{person}{Greg Hullender}.} \bibinfo{year}{2005}\natexlab{}.
\newblock \showarticletitle{Learning to rank using gradient descent}. In
  \bibinfo{booktitle}{\emph{Proc. of the 22nd International Conference on
  Machine Learning}}. \bibinfo{pages}{89--96}.
\newblock


\bibitem[\protect\citeauthoryear{Burges}{Burges}{2010}]%
        {burges2010ranknet}
\bibfield{author}{\bibinfo{person}{Christopher~J.C. Burges}.}
  \bibinfo{year}{2010}\natexlab{}.
\newblock \bibinfo{booktitle}{\emph{From {RankNet} to {LambdaRank} to
  {LambdaMART}: An Overview}}.
\newblock \bibinfo{type}{{T}echnical {R}eport} Technical Report MSR-TR-2010-82.
  \bibinfo{institution}{Microsoft Research}.
\newblock


\bibitem[\protect\citeauthoryear{Burges, Ragno, and Le}{Burges
  et~al\mbox{.}}{2006}]%
        {quoc2007learning}
\bibfield{author}{\bibinfo{person}{Christopher J.~C. Burges},
  \bibinfo{person}{Robert Ragno}, {and} \bibinfo{person}{Quoc~Viet Le}.}
  \bibinfo{year}{2006}\natexlab{}.
\newblock \showarticletitle{Learning to Rank with Nonsmooth Cost Functions}. In
  \bibinfo{booktitle}{\emph{Proc. of the 19th International Conference on
  Neural Information Processing Systems}}. \bibinfo{pages}{193--200}.
\newblock


\bibitem[\protect\citeauthoryear{Cao, Qin, Liu, Tsai, and Li}{Cao
  et~al\mbox{.}}{2007}]%
        {cao2007learning}
\bibfield{author}{\bibinfo{person}{Zhe Cao}, \bibinfo{person}{Tao Qin},
  \bibinfo{person}{Tie-Yan Liu}, \bibinfo{person}{Ming-Feng Tsai}, {and}
  \bibinfo{person}{Hang Li}.} \bibinfo{year}{2007}\natexlab{}.
\newblock \showarticletitle{Learning to rank: from pairwise approach to
  listwise approach}. In \bibinfo{booktitle}{\emph{Proc. of the 24th
  International Conference on Machine Learning}}. \bibinfo{pages}{129--136}.
\newblock


\bibitem[\protect\citeauthoryear{Carbonell and Goldstein}{Carbonell and
  Goldstein}{1998}]%
        {Carbonell+Goldstein:1998}
\bibfield{author}{\bibinfo{person}{Jaime Carbonell} {and} \bibinfo{person}{Jade
  Goldstein}.} \bibinfo{year}{1998}\natexlab{}.
\newblock \showarticletitle{The Use of MMR, Diversity-based Reranking for
  Reordering Documents and Producing Summaries}. In
  \bibinfo{booktitle}{\emph{Proc. of the 21st Annual International ACM SIGIR
  Conference on Research and Development in Information Retrieval}}.
  \bibinfo{pages}{335--336}.
\newblock


\bibitem[\protect\citeauthoryear{Chapelle and Chang}{Chapelle and
  Chang}{2011}]%
        {Chapelle+Chang:2011}
\bibfield{author}{\bibinfo{person}{O. Chapelle} {and} \bibinfo{person}{Y.
  Chang}.} \bibinfo{year}{2011}\natexlab{}.
\newblock \showarticletitle{Yahoo! Learning to Rank Challenge Overview}. In
  \bibinfo{booktitle}{\emph{Proc. of the Learning to Rank Challenge}}.
  \bibinfo{pages}{1--24}.
\newblock


\bibitem[\protect\citeauthoryear{Dehghani, Zamani, Severyn, Kamps, and
  Croft}{Dehghani et~al\mbox{.}}{2017}]%
        {Dehghani:SIGIR2017}
\bibfield{author}{\bibinfo{person}{Mostafa Dehghani}, \bibinfo{person}{Hamed
  Zamani}, \bibinfo{person}{Aliaksei Severyn}, \bibinfo{person}{Jaap Kamps},
  {and} \bibinfo{person}{W.~Bruce Croft}.} \bibinfo{year}{2017}\natexlab{}.
\newblock \showarticletitle{Neural Ranking Models with Weak Supervision}. In
  \bibinfo{booktitle}{\emph{Proc. of the 40th International ACM SIGIR
  Conference on Research and Development in Information Retrieval}}.
  \bibinfo{pages}{65--74}.
\newblock


\bibitem[\protect\citeauthoryear{Diaz}{Diaz}{2007}]%
        {diaz2007regularizing}
\bibfield{author}{\bibinfo{person}{Fernando Diaz}.}
  \bibinfo{year}{2007}\natexlab{}.
\newblock \showarticletitle{Regularizing query-based retrieval scores}.
\newblock \bibinfo{journal}{\emph{Information Retrieval}} \bibinfo{volume}{10},
  \bibinfo{number}{6} (\bibinfo{year}{2007}), \bibinfo{pages}{531--562}.
\newblock


\bibitem[\protect\citeauthoryear{Edizel, Mantrach, and Bai}{Edizel
  et~al\mbox{.}}{2017}]%
        {Edizel+al:2017}
\bibfield{author}{\bibinfo{person}{Bora Edizel}, \bibinfo{person}{Amin
  Mantrach}, {and} \bibinfo{person}{Xiao Bai}.}
  \bibinfo{year}{2017}\natexlab{}.
\newblock \showarticletitle{Deep Character-Level Click-Through Rate Prediction
  for Sponsored Search}. In \bibinfo{booktitle}{\emph{Proc. of the 40th
  International ACM SIGIR Conference on Research and Development in Information
  Retrieval}}. \bibinfo{pages}{305--314}.
\newblock


\bibitem[\protect\citeauthoryear{Friedman}{Friedman}{2001}]%
        {friedman2001greedy}
\bibfield{author}{\bibinfo{person}{Jerome~H Friedman}.}
  \bibinfo{year}{2001}\natexlab{}.
\newblock \showarticletitle{Greedy function approximation: a gradient boosting
  machine}.
\newblock \bibinfo{journal}{\emph{Annals of Statistics}} \bibinfo{volume}{29},
  \bibinfo{number}{5} (\bibinfo{year}{2001}), \bibinfo{pages}{1189--1232}.
\newblock


\bibitem[\protect\citeauthoryear{Gey}{Gey}{1994}]%
        {Gey:1994:IPR:188490.188560}
\bibfield{author}{\bibinfo{person}{Fredric~C. Gey}.}
  \bibinfo{year}{1994}\natexlab{}.
\newblock \showarticletitle{Inferring Probability of Relevance Using the Method
  of Logistic Regression}. In \bibinfo{booktitle}{\emph{Proc. of the 17th
  Annual International ACM SIGIR Conference on Research and Development in
  Information Retrieval}}. \bibinfo{pages}{222--231}.
\newblock
\showISBNx{0-387-19889-X}


\bibitem[\protect\citeauthoryear{Guo, Fan, Ai, and Croft}{Guo
  et~al\mbox{.}}{2016}]%
        {Guo+al:2016}
\bibfield{author}{\bibinfo{person}{Jiafeng Guo}, \bibinfo{person}{Yixing Fan},
  \bibinfo{person}{Qingyao Ai}, {and} \bibinfo{person}{W.~Bruce Croft}.}
  \bibinfo{year}{2016}\natexlab{}.
\newblock \showarticletitle{A Deep Relevance Matching Model for Ad-hoc
  Retrieval}. In \bibinfo{booktitle}{\emph{Proc. of the 25rd ACM International
  Conference on Information and Knowledge Management}}.
  \bibinfo{pages}{55--64}.
\newblock


\bibitem[\protect\citeauthoryear{Huang, He, Gao, Deng, Acero, and Heck}{Huang
  et~al\mbox{.}}{2013}]%
        {Huang+al:2013}
\bibfield{author}{\bibinfo{person}{Po-Sen Huang}, \bibinfo{person}{Xiaodong
  He}, \bibinfo{person}{Jianfeng Gao}, \bibinfo{person}{Li Deng},
  \bibinfo{person}{Alex Acero}, {and} \bibinfo{person}{Larry Heck}.}
  \bibinfo{year}{2013}\natexlab{}.
\newblock \showarticletitle{Learning Deep Structured Semantic Models for Web
  Search Using Clickthrough Data}. In \bibinfo{booktitle}{\emph{Proc. of the
  22nd ACM International Conference on Information and Knowledge Management}}.
  \bibinfo{pages}{2333--2338}.
\newblock


\bibitem[\protect\citeauthoryear{J{\"a}rvelin and
  Kek{\"a}l{\"a}inen}{J{\"a}rvelin and Kek{\"a}l{\"a}inen}{2002}]%
        {jarvelin2002cumulated}
\bibfield{author}{\bibinfo{person}{Kalervo J{\"a}rvelin} {and}
  \bibinfo{person}{Jaana Kek{\"a}l{\"a}inen}.} \bibinfo{year}{2002}\natexlab{}.
\newblock \showarticletitle{Cumulated gain-based evaluation of IR techniques}.
\newblock \bibinfo{journal}{\emph{ACM Transactions on Information Systems}}
  \bibinfo{volume}{20}, \bibinfo{number}{4} (\bibinfo{year}{2002}),
  \bibinfo{pages}{422--446}.
\newblock


\bibitem[\protect\citeauthoryear{Jiang, Wen, Dou, Zhao, Nie, and Yue}{Jiang
  et~al\mbox{.}}{2017}]%
        {Jiang+al:2017}
\bibfield{author}{\bibinfo{person}{Zhengbao Jiang}, \bibinfo{person}{Ji-Rong
  Wen}, \bibinfo{person}{Zhicheng Dou}, \bibinfo{person}{Wayne~Xin Zhao},
  \bibinfo{person}{Jian-Yun Nie}, {and} \bibinfo{person}{Ming Yue}.}
  \bibinfo{year}{2017}\natexlab{}.
\newblock \showarticletitle{Learning to Diversify Search Results via Subtopic
  Attention}. In \bibinfo{booktitle}{\emph{Proc. of the 40th International ACM
  SIGIR Conference on Research and Development in Information Retrieval}}.
  \bibinfo{pages}{545--554}.
\newblock


\bibitem[\protect\citeauthoryear{Joachims}{Joachims}{2002}]%
        {Joachims:2002}
\bibfield{author}{\bibinfo{person}{Thorsten Joachims}.}
  \bibinfo{year}{2002}\natexlab{}.
\newblock \showarticletitle{Optimizing Search Engines Using Clickthrough Data}.
  In \bibinfo{booktitle}{\emph{Proc. of the 8th ACM SIGKDD International
  Conference on Knowledge Discovery and Data Mining}}.
  \bibinfo{pages}{133--142}.
\newblock


\bibitem[\protect\citeauthoryear{Joachims}{Joachims}{2006}]%
        {joachims2006training}
\bibfield{author}{\bibinfo{person}{Thorsten Joachims}.}
  \bibinfo{year}{2006}\natexlab{}.
\newblock \showarticletitle{Training linear SVMs in linear time}. In
  \bibinfo{booktitle}{\emph{Proc. of the 12th ACM SIGKDD International
  Conference on Knowledge Discovery and Data Mining}}.
  \bibinfo{pages}{217--226}.
\newblock


\bibitem[\protect\citeauthoryear{Joachims, Granka, Pan, Hembrooke, and
  Gay}{Joachims et~al\mbox{.}}{2005}]%
        {Joachims+al:2005}
\bibfield{author}{\bibinfo{person}{Thorsten Joachims}, \bibinfo{person}{Laura
  Granka}, \bibinfo{person}{Bing Pan}, \bibinfo{person}{Helene Hembrooke},
  {and} \bibinfo{person}{Geri Gay}.} \bibinfo{year}{2005}\natexlab{}.
\newblock \showarticletitle{Accurately Interpreting Clickthrough Data As
  Implicit Feedback}. In \bibinfo{booktitle}{\emph{Proc. of the 28th Annual
  International ACM SIGIR Conference on Research and Development in Information
  Retrieval}}. \bibinfo{pages}{154--161}.
\newblock


\bibitem[\protect\citeauthoryear{Joachims, Swaminathan, and Schnabel}{Joachims
  et~al\mbox{.}}{2017}]%
        {Joachims:WSDM17}
\bibfield{author}{\bibinfo{person}{Thorsten Joachims}, \bibinfo{person}{Adith
  Swaminathan}, {and} \bibinfo{person}{Tobias Schnabel}.}
  \bibinfo{year}{2017}\natexlab{}.
\newblock \showarticletitle{Unbiased Learning-to-Rank with Biased Feedback}. In
  \bibinfo{booktitle}{\emph{Proc. of the 10th ACM International Conference on
  Web Search and Data Mining}}. \bibinfo{pages}{781--789}.
\newblock


\bibitem[\protect\citeauthoryear{Liu}{Liu}{2009}]%
        {liu2009learning}
\bibfield{author}{\bibinfo{person}{Tie-Yan Liu}.}
  \bibinfo{year}{2009}\natexlab{}.
\newblock \showarticletitle{Learning to rank for information retrieval}.
\newblock \bibinfo{journal}{\emph{Foundations and Trends in Information
  Retrieval}} \bibinfo{volume}{3}, \bibinfo{number}{3} (\bibinfo{year}{2009}),
  \bibinfo{pages}{225--331}.
\newblock


\bibitem[\protect\citeauthoryear{Manning, Raghavan, and Sch\"{u}tze}{Manning
  et~al\mbox{.}}{2008}]%
        {Manning:2008:IIR:1394399}
\bibfield{author}{\bibinfo{person}{Christopher~D. Manning},
  \bibinfo{person}{Prabhakar Raghavan}, {and} \bibinfo{person}{Hinrich
  Sch\"{u}tze}.} \bibinfo{year}{2008}\natexlab{}.
\newblock \bibinfo{booktitle}{\emph{Introduction to Information Retrieval}}.
\newblock \bibinfo{publisher}{Cambridge University Press},
  \bibinfo{address}{New York, NY, USA}.
\newblock
\showISBNx{0521865719, 9780521865715}


\bibitem[\protect\citeauthoryear{Mitra, Diaz, and Craswell}{Mitra
  et~al\mbox{.}}{2017}]%
        {Mitra+al:2017}
\bibfield{author}{\bibinfo{person}{Bhaskar Mitra}, \bibinfo{person}{Fernando
  Diaz}, {and} \bibinfo{person}{Nick Craswell}.}
  \bibinfo{year}{2017}\natexlab{}.
\newblock \showarticletitle{Learning to Match Using Local and Distributed
  Representations of Text for Web Search}. In \bibinfo{booktitle}{\emph{Proc.
  of the 26th International Conference on World Wide Web}}.
  \bibinfo{pages}{1291--1299}.
\newblock


\bibitem[\protect\citeauthoryear{Pang, Lan, Guo, Xu, Xu, and Cheng}{Pang
  et~al\mbox{.}}{2017}]%
        {DeepRank:2017}
\bibfield{author}{\bibinfo{person}{Liang Pang}, \bibinfo{person}{Yanyan Lan},
  \bibinfo{person}{Jiafeng Guo}, \bibinfo{person}{Jun Xu},
  \bibinfo{person}{Jingfang Xu}, {and} \bibinfo{person}{Xueqi Cheng}.}
  \bibinfo{year}{2017}\natexlab{}.
\newblock \showarticletitle{DeepRank: A New Deep Architecture for Relevance
  Ranking in Information Retrieval}. In \bibinfo{booktitle}{\emph{Proc. of the
  2017 ACM Conference on Information and Knowledge Management}}.
  \bibinfo{pages}{257--266}.
\newblock


\bibitem[\protect\citeauthoryear{Pasumarthi, Wang, Li, Bruch, Bendersky,
  Najork, Pfeifer, Golbandi, Anil, and Wolf}{Pasumarthi et~al\mbox{.}}{2018}]%
        {TensorflowRanking2018}
\bibfield{author}{\bibinfo{person}{Rama~Kumar Pasumarthi},
  \bibinfo{person}{Xuanhui Wang}, \bibinfo{person}{Cheng Li},
  \bibinfo{person}{Sebastian Bruch}, \bibinfo{person}{Michael Bendersky},
  \bibinfo{person}{Marc Najork}, \bibinfo{person}{Jan Pfeifer},
  \bibinfo{person}{Nadav Golbandi}, \bibinfo{person}{Rohan Anil}, {and}
  \bibinfo{person}{Stephan Wolf}.} \bibinfo{year}{2018}\natexlab{}.
\newblock \bibinfo{title}{TF-Ranking: Scalable TensorFlow Library for
  Learning-to-Rank}.
\newblock   (\bibinfo{year}{2018}).
\newblock
\showeprint{arXiv:1812.00073}


\bibitem[\protect\citeauthoryear{Qin and Liu}{Qin and Liu}{2013}]%
        {DBLP:journals/corr/QinL13}
\bibfield{author}{\bibinfo{person}{Tao Qin} {and} \bibinfo{person}{Tie-Yan
  Liu}.} \bibinfo{year}{2013}\natexlab{}.
\newblock \bibinfo{title}{Introducing {LETOR} 4.0 Datasets}.
\newblock   (\bibinfo{year}{2013}).
\newblock
\showeprint[arxiv]{1306.2597}


\bibitem[\protect\citeauthoryear{Qin, Liu, Zhang, Wang, and Li}{Qin
  et~al\mbox{.}}{2008}]%
        {qin2009global}
\bibfield{author}{\bibinfo{person}{Tao Qin}, \bibinfo{person}{Tie-Yan Liu},
  \bibinfo{person}{Xu-Dong Zhang}, \bibinfo{person}{De-Sheng Wang}, {and}
  \bibinfo{person}{Hang Li}.} \bibinfo{year}{2008}\natexlab{}.
\newblock \showarticletitle{Global ranking using continuous conditional random
  fields}. In \bibinfo{booktitle}{\emph{Proc. of the 21st International
  Conference on Neural Information Processing Systems}}.
  \bibinfo{pages}{1281--1288}.
\newblock


\bibitem[\protect\citeauthoryear{Robert and Casella}{Robert and
  Casella}{2005}]%
        {MonteCarlo}
\bibfield{author}{\bibinfo{person}{Christian~P. Robert} {and}
  \bibinfo{person}{George Casella}.} \bibinfo{year}{2005}\natexlab{}.
\newblock \bibinfo{booktitle}{\emph{Monte Carlo Statistical Methods}}.
\newblock \bibinfo{publisher}{Springer-Verlag}.
\newblock


\bibitem[\protect\citeauthoryear{Taylor, Guiver, Robertson, and Minka}{Taylor
  et~al\mbox{.}}{2008}]%
        {Taylor+al:2008}
\bibfield{author}{\bibinfo{person}{Michael Taylor}, \bibinfo{person}{John
  Guiver}, \bibinfo{person}{Stephen Robertson}, {and} \bibinfo{person}{Tom
  Minka}.} \bibinfo{year}{2008}\natexlab{}.
\newblock \showarticletitle{SoftRank: Optimizing Non-smooth Rank Metrics}. In
  \bibinfo{booktitle}{\emph{Proc. of the 1st International Conference on Web
  Search and Data Mining}}. \bibinfo{pages}{77--86}.
\newblock


\bibitem[\protect\citeauthoryear{Wang, Bendersky, Metzler, and Najork}{Wang
  et~al\mbox{.}}{2016}]%
        {Wang+al:2016}
\bibfield{author}{\bibinfo{person}{Xuanhui Wang}, \bibinfo{person}{Michael
  Bendersky}, \bibinfo{person}{Donald Metzler}, {and} \bibinfo{person}{Marc
  Najork}.} \bibinfo{year}{2016}\natexlab{}.
\newblock \showarticletitle{Learning to Rank with Selection Bias in Personal
  Search}. In \bibinfo{booktitle}{\emph{Proc. of the 39th International ACM
  SIGIR Conference on Research and Development in Information Retrieval}}.
  \bibinfo{pages}{115--124}.
\newblock


\bibitem[\protect\citeauthoryear{Wang, Golbandi, Bendersky, Metzler, and
  Najork}{Wang et~al\mbox{.}}{2018}]%
        {Wang+al:2018}
\bibfield{author}{\bibinfo{person}{Xuanhui Wang}, \bibinfo{person}{Nadav
  Golbandi}, \bibinfo{person}{Michael Bendersky}, \bibinfo{person}{Donald
  Metzler}, {and} \bibinfo{person}{Marc Najork}.}
  \bibinfo{year}{2018}\natexlab{}.
\newblock \showarticletitle{Position Bias Estimation for Unbiased Learning to
  Rank in Personal Search}. In \bibinfo{booktitle}{\emph{Proc. of the 11th
  International Conference on Web Search and Data Mining}}. \bibinfo{pages}{610
  --618}.
\newblock


\bibitem[\protect\citeauthoryear{Xia, Liu, Wang, Zhang, and Li}{Xia
  et~al\mbox{.}}{2008}]%
        {xia2008listwise}
\bibfield{author}{\bibinfo{person}{Fen Xia}, \bibinfo{person}{Tie-Yan Liu},
  \bibinfo{person}{Jue Wang}, \bibinfo{person}{Wensheng Zhang}, {and}
  \bibinfo{person}{Hang Li}.} \bibinfo{year}{2008}\natexlab{}.
\newblock \showarticletitle{Listwise approach to learning to rank: theory and
  algorithm}. In \bibinfo{booktitle}{\emph{Proc. of the 25th International
  Conference on Machine Learning}}. \bibinfo{pages}{1192--1199}.
\newblock


\bibitem[\protect\citeauthoryear{Xia, Xu, Lan, Guo, and Cheng}{Xia
  et~al\mbox{.}}{2016}]%
        {Xia+al:2016}
\bibfield{author}{\bibinfo{person}{Long Xia}, \bibinfo{person}{Jun Xu},
  \bibinfo{person}{Yanyan Lan}, \bibinfo{person}{Jiafeng Guo}, {and}
  \bibinfo{person}{Xueqi Cheng}.} \bibinfo{year}{2016}\natexlab{}.
\newblock \showarticletitle{Modeling Document Novelty with Neural Tensor
  Network for Search Result Diversification}. In
  \bibinfo{booktitle}{\emph{Proc. of the 39th International ACM SIGIR
  Conference on Research and Development in Information Retrieval}}.
  \bibinfo{pages}{395--404}.
\newblock


\bibitem[\protect\citeauthoryear{Xu and Li}{Xu and Li}{2007}]%
        {Jun+Hang:2007}
\bibfield{author}{\bibinfo{person}{Jun Xu} {and} \bibinfo{person}{Hang Li}.}
  \bibinfo{year}{2007}\natexlab{}.
\newblock \showarticletitle{AdaRank: A Boosting Algorithm for Information
  Retrieval}. In \bibinfo{booktitle}{\emph{Proc. of the 30th Annual
  International ACM SIGIR Conference on Research and Development in Information
  Retrieval}}. \bibinfo{pages}{391--398}.
\newblock


\bibitem[\protect\citeauthoryear{Ye and Doermann}{Ye and Doermann}{2013}]%
        {ye2013combining}
\bibfield{author}{\bibinfo{person}{Peng Ye} {and} \bibinfo{person}{David
  Doermann}.} \bibinfo{year}{2013}\natexlab{}.
\newblock \showarticletitle{Combining preference and absolute judgements in a
  crowd-sourced setting}. In \bibinfo{booktitle}{\emph{ICML 2013 Workshop on
  Machine Learning Meets Crowdsourcing}}.
\newblock


\bibitem[\protect\citeauthoryear{Zamani, Bendersky, Wang, and Zhang}{Zamani
  et~al\mbox{.}}{2017}]%
        {Zamani+al:2017}
\bibfield{author}{\bibinfo{person}{Hamed Zamani}, \bibinfo{person}{Michael
  Bendersky}, \bibinfo{person}{Xuanhui Wang}, {and} \bibinfo{person}{Mingyang
  Zhang}.} \bibinfo{year}{2017}\natexlab{}.
\newblock \showarticletitle{Situational Context for Ranking in Personal
  Search}. In \bibinfo{booktitle}{\emph{Proc. of the 26th International
  Conference on World Wide Web}}. \bibinfo{pages}{1531--1540}.
\newblock


\end{thebibliography}

\end{document}